\documentclass[]{article}

\usepackage{arxiv}

\usepackage[utf8]{inputenc} 
\usepackage[T1]{fontenc}    
\usepackage{hyperref}       
\usepackage{url}            
\usepackage{booktabs}       
\usepackage{amsfonts}       
\usepackage{nicefrac}       
\usepackage{microtype}      
\usepackage{graphicx}
\usepackage[numbers]{natbib}
\usepackage{doi}
\usepackage{arydshln}
\usepackage{amsmath}

\title{Randomization-based Inference for MCP-Mod}

\author{ \href{https://orcid.org/0009-0003-9512-681X}{\includegraphics[scale=0.06]{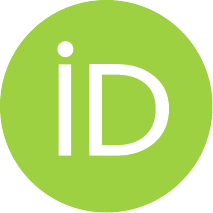}\hspace{1mm}Lukas Pin}\\
	MRC Biostatistics Unit \\
	University of Cambridge\\
	Cambridge, UK \\
	\texttt{lukas.pin@mrc-bsu.cam.ac.uk} \\
	\And
	\href{https://orcid.org/0000-0002-1626-2588}{\includegraphics[scale=0.06]{orcid.pdf}\hspace{1mm}Oleksandr Sverdlov} \\
	Early Development Analytics\\
	Novartis Pharmaceuticals Corporation\\
	East Hanover, NJ, USA \\
	\texttt{oleksandr.sverdlov@novartis.com} \\
	\And
	\href{https://orcid.org/0000-0002-2008-8340}{\includegraphics[scale=0.06]{orcid.pdf}\hspace{1mm}Frank Bretz} \\
	Advanced Methodology and Data Science\\
	Novartis Pharma AG\\
	Basel, Switzerland \\
	\texttt{frank.bretz@novartis.com} \\
	\And
	\href{https://orcid.org/0000-0002-6294-8185}{\includegraphics[scale=0.06]{orcid.pdf}\hspace{1mm}Björn Bornkamp} \\
	Advanced Methodology and Data Science\\
	Novartis Pharma AG\\
	Basel, Switzerland \\
	\texttt{bjorn.bornkamp@novartis.com} \\
}

\renewcommand{\shorttitle}{\textit{Randomization-based Inference for MCP-Mod}}

\hypersetup{
pdftitle={Randomization-based Inference for MCP-Mod},
pdfsubject={biostatistics, dose-finding trials},
pdfauthor={Lukas Pin, Oleksandr Sverdlov, Frank Bretz, Björn Bornkamp},
pdfkeywords={Randomization test, Dose finding, Multiple testing, Finite sample inference, Time trends, Penalized maximum likelihood estimation},
}

\begin{document}
\maketitle

\begin{abstract}
	Dose selection is critical in pharmaceutical drug development, as it directly impacts therapeutic efficacy and patient safety of a drug. The Generalized Multiple Comparison Procedures and Modeling (MCP-Mod) approach is commonly used in Phase II trials for testing and estimation of dose-response relationships. However, its effectiveness in small sample sizes, particularly with binary endpoints, is hindered by issues like complete separation in logistic regression, leading to non-existence of estimates. Motivated by an actual clinical trial using the MCP-Mod approach, this paper introduces penalized maximum likelihood estimation (MLE) and randomization-based inference techniques to address these challenges. Randomization-based inference allows for exact finite sample inference, while population-based inference for MCP-Mod typically relies on asymptotic approximations. Simulation studies demonstrate that randomization-based tests can enhance statistical power in small to medium-sized samples while maintaining control over type-I error rates, even in the presence of time trends. Our results show that residual-based randomization tests using penalized MLEs not only improve computational efficiency but also outperform standard randomization-based methods, making them an adequate choice for dose-finding analyses within the MCP-Mod framework. Additionally, we apply these methods to pharmacometric settings, demonstrating their effectiveness in such scenarios. The results in this paper underscore the potential of randomization-based inference for the analysis of dose-finding trials, particularly in small sample contexts.
\end{abstract}


\section{Introduction}\label{sec1}
Identifying an adequate dose is an important objective of pharmaceutical drug development, directly impacting both therapeutic efficacy and patient safety considered for market authorization. The development process of a pharmaceutical drug involves several critical phases: Phase I aims to identify the maximum tolerated dose \citep{neue:bran:gspo:2008}, while Phase II focuses on modeling dose-response and dose-exposure-response relationships \citep{hsu:2009} and selecting promising dose levels for Phase III. If efficacy and benefit-risk are confirmed in Phase III, the drug is submitted to regulatory agencies for market authorization. Post-marketing studies may further refine the understanding of dose-response relationships across diverse subgroups, defined by factors such as gender, age, disease severity, and other covariates \citep{rube:1995a, rube:1995b, iche4:1994}.
Choosing the correct dose level is crucial: a dose too low may render the drug ineffective, while a dose too high could result in unacceptable toxicities. Therefore, precise dose selection is vital to optimize therapeutic outcomes and minimize adverse effects, ultimately contributing to the clinical development program's overall success.

One popular method for dose selection in Phase II trials is the Multiple Comparison Procedures and Modeling (MCP-Mod) approach introduced by Bretz et al. (2005) \cite{bret:pinh:bran:2005}. 
The MCP-Mod approach models the entire dose-response relationship while accounting for model uncertainty. Originally designed for normal, homoscedastic, and independent data, the methodology was later expanded by Pinheiro et al. (2014) \cite{pinh:born:glim:2014} to the generalized MCP-Mod approach to encompass binary, count, and time-to-event data. This expansion allows for the inclusion of generalized nonlinear models, linear and nonlinear mixed effects models, and Cox proportional hazards models, thereby integrating covariates into the analysis; see Section \ref{Sec:MCP-Mod} for more details.

Since its development, the generalized MCP-Mod has been applied to various Phase II studies \citep{maurer2019ligelizumab, maurer2022remibrutinib, bowman2022safety, zhang2024iptacopan}. In trials with smaller sample size, however, issues related to inference and estimation can arise, particularly in trials with a binary endpoint.
Generalized MCP-Mod utilizes population-based inference and typically relies on on asymptotic properties for power and type-I error rates \citep{pinh:born:glim:2014}. However, for small sample size these approximations may not be accurate. Additionally, e.g. in trials with slow recruitment, potential time trends can pose significant challenges. Another concern is the requirement for the existence of maximum likelihood estimates (MLEs) for testing, which is not always guaranteed.

The paper is structured as follows: 
In Section \ref{Sec:Trial} we provide a trial example. 
In Section \ref{Sec:Estimation} we explain how complete separation may lead to non-existence of maximum likelihood estimates for binary data.
To address this issue, we propose the use of penalized maximum likelihood estimation.
Moreover, we use randomization-based inference methods, detailed in Sections \ref{Sec:Estimation} and \ref{Sec:RandBasedInf}, respectively to address the issues concerning asymptotic test properties and time trends. These methods lead to several novel approaches that we compare in two simulation studies in Section \ref{Sec:Simulations}. The first simulation study, which is presented in Section \ref{Sec:MainSimulation}, compares the different methods across several clinical trial settings that are adapted from the example that is introduced in Section \ref{Sec:Trial}. The second simulation study demonstrates how randomization-based inference can be applied for continuous outcomes with simulation scenarios motivated by a pharmacometric case study \citep{buatois2021clrt} in Section \ref{Sec:Pharmaco}. This is followed by a discussion and suggestions for future research directions in Section \ref{Sec:Discussion}.

\section{A Phase II Dose Finding Trial}\label{Sec:Trial}

In this section, we introduce a clinical trial example that is motivated from a real trial. For the purpose of presentation the example was modified from the real trial to maintain confidentiality and simplify presentation. The adapted example serves as one scenario for the simulation study in Section \ref{Sec:Simulations}. 

The trial is a randomized, double-blind investigation. In this paper we assume that the trial is designed to assess the efficacy and safety of a drug across three different dose levels: 10 mg, 25 mg, and 100 mg, compared to a placebo (0 mg). We have in total $n = 49$ patients, which were randomized using a $1:2:2:2$ randomization ratio (placebo against the three active doses). For simplicity and the purpose here we assume permuted block randomization with blocks of fixed length of $B=7$ would be employed to sequentially perform treatment assignments and to ensure that 7 patients were allocated to the placebo group and 14 patients to each of the active doses.
The primary endpoint of the trial is binary, where a value of $1$ indicates a successful treatment outcome and a value of $0$ treatment failure. The trial design aims to achieve $80$\% power to detect the presence of dose-response signal at a one-sided significance level of $10$\% using the generalized MCP-Mod approach.
The assumed response rates are $20$\% for the placebo group and $80$\% for the highest dose group (100 mg). These parameters provide the foundation for the simulation study in Section \ref{Sec:Simulations}. In the next Section we discuss and introduce methodology to analyze this trial example.

\section{Methodology}

In this section, we briefly introduce the generalized MCP-Mod approach and address the challenges of estimation with binary outcomes in small sample sizes. To overcome these challenges, we adapt the MCP-Mod framework by incorporating two key methods: Firth's method to handle complete separation and randomization-based inference to ensure a robust and valid inference framework. The latter provides appropriate type-I error rate control and enhances robustness against time trends, making it well-suited for small-sample settings.


Throughout this section
we consider a parallel group, randomized, placebo-controlled trial, where \( n \) patients are randomized among doses \( d_0, \ldots, d_{k-1} \), with \( d_0 \) representing the placebo dose. The treatment assignment for patient \( i \) is indicated by \( Z_i \), which can take any value from the set \(\{d_0, \ldots, d_{k-1}\}\). Collectively, the treatment assignments are represented by the vector \(\mathbf{Z} = (Z_1, \ldots, Z_n)\), which is drawn from the randomization probability distribution \(\mathcal{P}(\mathbf{Z})\) defined over all possible treatment sequences. 
Each patient \( i \) has $k$ potential outcomes \( y_{i,z} \) corresponding to each treatment \( z \) within \(\{d_0, \ldots, d_{k-1}\}\). Additionally, patient \( i \) has a covariate
\(\mathbf{x}_{i}\).
Upon realization of the treatment sequence \(\mathbf{z}\), only the outcome \( y_{i,z_i} \) is observed for each patient \( i \) and assigned treatment \( z_i \); all other potential outcomes for that patient remain unobserved.

\subsection{Generalized MCP-Mod}\label{Sec:MCP-Mod}

As previously mentioned, the generalized MCP-Mod approach can be used for testing and modeling the relationship between different doses and their responses while accommodating uncertainty in the dose-response models.
This methodology builds upon a well-established framework used in clinical Phase II dose-response studies for normally distributed, homoscedastic responses measured at a single time point in parallel group designs \citep{bret:pinh:bran:2005}. The generalized MCP-Mod expands on this by addressing more complex scenarios, such as binary outcomes, count data or time-to-event endpoints \citep{pinh:born:glim:2014}. An implementation of generalized MCP-Mod is publicly available in form of the \texttt{DoseFinding}-package in \texttt{R} \citep{dosefinding2024}.
Two main steps characterize the MCP-Mod approach:
\begin{enumerate}
    \item \textbf{Testing a dose-response signal:} The first step involves using a multiple contrast test to evaluate whether there is significant evidence against the null hypothesis 
    \begin{equation}
        \label{eq:H0Population}
          H_0: \mu_{0}=\ldots=\mu_{k-1}
    \end{equation}
    that all treatment groups have the same mean response. This test employs contrast coefficients optimized for detecting specific dose-response shapes, as defined by pre-specified candidate models, and controls the type-I error rate at a designated significance level. 
    \vspace{0.5cm}
    \item \textbf{Estimating a dose-response relationship:} If the null hypothesis $H_0$ is rejected, the second step involves modeling the dose-response relationship. This step frequently uses model averaging over all candidate dose-response models to estimate the relationship more robustly \citep{schorning2016model}.
\end{enumerate}

The MCP-Mod approach is recognized by both the EMA \cite{EMA2024} and the FDA \cite{FDA2024} as part of their qualification and fit-for-purpose initiatives, highlighting its effectiveness as a statistical methodology for model-based design and analysis in Phase II dose-finding studies. In this paper, we retain the core modeling framework of MCP-Mod but focus on refining the testing component of the method by utilizing a randomization-based inference framework instead of the population-based framework that the methodology was originally proposed in.

\subsection{Challenges in Estimation with Binary Outcomes for Small Samples}\label{Sec:Estimation}

In this section, we discuss complete or quasicomplete separation in logistic regression, which leads to non-existence of unique MLEs \citep{silvapulle1981, albert1984existence}.
Complete separation in binary classification models refers to a situation where the predictor variables perfectly predict the outcome variable. In other words, there exists a hyperplane that can separate the two classes without any errors. Meaning that if there is a combination of predictor variables such that all observations with value $1$ lie on one side of the hyperplane and all observations with value $0$ lie on the other side, then complete separation is achieved.
Quasicomplete separation is a related scenario where some data points may lie exactly on the hyperplane, which also leads to non-existence of MLEs.

Heinze et al. (2020) \cite{heinze2002solution} propose using Firth's method \citep{firth1993bias}, which modifies the score function of logistic regression to reduce the bias of MLEs, which tend to be biased away from zero. This approach ensures finite estimates closer to zero. Specifically, estimates using Firth's method always exist, even in situation of complete or quasicomplete separation. The estimates can be calculated efficiently using the \texttt{logistf}-package in \texttt{R} \cite{logistf2023} or \texttt{PROC LOGISTIC} in \texttt{SAS} \cite{SASLogistic}.

Complete separation is more likely to occur for smaller sample sizes. This phenomenon is evident in our trial example from Section \ref{Sec:Trial}, as simulated in Section \ref{Sec:MainSimulation}. When we increase the sample size under the null hypothesis, the likelihood of complete separation decreases significantly. Based on 10,000 simulations, we observed that the likelihood of complete separation is 18.02\% with a sample size of 49. This likelihood decreases to 3.36\% when the sample size increases to 98 and no cases of complete separation are observed with a sample size of 490. These findings indicate that complete separation is much more common in smaller sample sizes and becomes exceedingly rare as the sample size grows.

\subsection{Randomization-based Inference}\label{Sec:RandBasedInf}
In this section, we develop randomization tests \citep{imbens2015causal,rosenberger2016randomization} for the generalized MCP-Mod approach. We first explain how the inference framework of randomization tests differs from the framework of population-based tests. Next, we introduce suitable test statistics for randomization-based inference that are similar or identical to those employed in the MCP-Mod approach within population sampling framework. Finally, we introduce different randomization procedures and discuss how they influence the randomization tests.

\subsubsection{Population-based versus Randomization-based Inference}
In population-based inference, we operate under the premise that our trial sample is drawn from a larger super-population. The trial participants are assumed to constitute a random sample from this super-population. Thus, in this framework, we infer characteristics of the larger super-population based on the observed outcomes $y_{i,z_i}$ for $i=1,\ldots,n$.
When employing a population-based model, we specify a statistical model that describes how the data are generated from the population and articulate a hypothesis we want to test. For methods such as MCP-Mod, this involves hypothesis \eqref{eq:H0Population} outlined in Section \ref{Sec:MCP-Mod}. Based on the observed data, the relevant parameters are estimated and hypothesis tests are conducted to determine whether there are statistically significant differences between treatment groups. The results are interpreted within the context of the larger super-population.
By adopting a population-based perspective, researchers aim to generalize the effects of treatments or interventions beyond the confines of the trial sample to a wider demographic. The model depends on the assumption that we independently sample from a super-population, which is strong and unverifiable \citep{msaouel2023role}. Furthermore, parametric distributional assumptions are often made. If these assumptions are violated, it can lead to biased estimates and incorrect inferences. 

In contrast, in randomization-based inference, we focus on the finite trial sample without making assumptions about independent sampling from a larger population. 
We then investigate, for example, the strong null hypothesis 
\begin{equation}
\label{eqn:h0}
  H_0: y_{i,d_0}=\ldots=y_{i,d_{k-1}}\; \quad \text{for all } \quad  i \in \{1,...,n\}.
\end{equation}
For testing hypothesis (\ref{eqn:h0}), a test-statistic $S(\mathbf{Z})$ is used for which large values of the test-statistic indicate evidence against the null-hypothesis. 
The hypothesis \eqref{eqn:h0} allows to investigate the distribution of $S(\mathbf{Z})$ under the randomization probability distribution $\mathcal{P}$, because it implies that all potential outcomes of patient $i$ are the same under all treatments. The test statistic $S(\mathbf{Z})$ can be calculated for any allocation sequence $\mathbf{Z}$ without knowing the potential outcome of a patient a treatment to which they were not assigned, see Section \ref{Sec:GenervalvsResidu} and \ref{Sec:RandProcedures}.
The exact p-value for the observed sequence $\mathbf{Z}_{obs}$ can be calculated as $\sum_{s} \pi_s \textbf{1}(S(\mathbf{Z}_s) \geq S(\mathbf{Z}_{obs}))$, where the sum is taken over all possible randomization sequences in the reference set, denoted as $\mathcal{R} $. Here, $\pi_s$ represents the probability of sequence $s$ occurring, and $\textbf{1}(\cdot)$ is the indicator function.
In practice, enumerating the complete reference set will be computationally challenging, but the p-value can be approximated by Monte Carlo sampling from $\mathcal{P}$, i.e., by generating independent realizations $\mathbf{Z}_1,\ldots,\mathbf{Z}_P$ from $\mathcal{P}$.  A p-value for the null hypothesis \eqref{eqn:h0} can then be approximated based on $\frac{1}{n_{rand}}\sum_{l=1}^{n_{rand}}\textbf{1}(S(\mathbf{Z}_l) \geq S(\mathbf{Z}_{obs}))$, where $n_{rand}$ is some large number. Plamadeala et al. (2012) \cite{Plamadeala2012} and Zhang et al. (2023) \cite{Zhang2023} provide some formal approaches for justifying the choice of $n_{rand}$ in practice. Note that the inference does not rely on any distributional assumptions on the outcome, only the fact that the treatment indicator was randomized is utilized.


\subsubsection{Randomization-based inference for MCP-Mod}\label{Sec:GenervalvsResidu}

In the following, we introduce two possible test statistics for randomization-based inference that are similar or identical to those employed in the original generalized MCP-Mod approach. 

\textbf{Test statistic 1: Generalized MCP-Mod}\\
Let $\mathbf{D}_{\mathbf{Z}}$ be the $(n \times k)$ matrix that has entries $d_{i,j}=1$ if $Z_i=d_j$ and $0$ otherwise. Furthermore, let $\mathbf{d}(\mathbf{Z})_i=(d_{i,0}, \ldots, d_{i,k-1})$.
A generalized linear model with linear predictor $\eta_i=  \mathbf{d}(\mathbf{Z})_i'\boldsymbol{\delta} + \mathbf{x}_i'\boldsymbol{\beta}$ is fitted, where  $\boldsymbol{\delta}=(\delta_0,...,\delta_{k-1})'$ are the dose specific intercepts. 
Based on fitting this model (using either likelihood or penalized likelihood estimation), population-average model predictions can be performed based on the obtained parameter estimates. That is, the population average estimate for dose $j$ on the linear predictor scale would be $\delta_j+\frac{1}{n}\sum_{i=1}^n\mathbf{x}_i'\boldsymbol{\beta}$, 
providing an estimate of the group means $\widehat{\boldsymbol{\mu}}_{\mathbf{Z}}$ with covariance matrix $\mathbf{S}_{\mathbf{Z}}$ calculated under the assumptions of the used model. The test statistic is then given by\begin{eqnarray}\label{eq:MCP}
  S_1(\mathbf{Z}) =\underset{m\in\{1,...,M\}}{\max} T_{1m}(\mathbf{Z}), \; \mathrm{where} \; T_{1m}(\mathbf{Z}) = \frac{\mathbf{c}_m'\widehat{\boldsymbol{\mu}}_{\mathbf{Z}}}{\sqrt{\mathbf{c}_m'\mathbf{S}_{\mathbf{Z}}\mathbf{c}_m}},
\end{eqnarray}
where $\mathbf{c}_1,...\mathbf{c}_M$ are the optimal contrasts underlying the $M$ candidate model shapes, e.g. see Figure \ref{fig:CandModels}. For more information on how the contrasts are calculated see Pinheiro et al. (2014) \cite{pinh:born:glim:2014}.

\textbf{Test statistic 2: Residual-based inference}\\
One disadvantage of the previous approach is that for each Monte Carlo sample $\mathbf{Z}_1,\ldots,\mathbf{Z}_P$ the underlying generalized linear model would need to be fitted which is why we consider an approach motivated by Parhat et al. (2014) \cite{parhat2014conditional} that avoids those calculations.
Instead, we can fit a generalized linear model with linear predictor $\eta_i=\alpha + \mathbf{x}_i'\boldsymbol{\beta}$ once. Second, we extract the residuals $r_i=y_i - g(\hat{\eta}_i)=y_i - g(\hat{\alpha}+\mathbf{x}_i'\widehat{\boldsymbol{\beta}})$ from this model, denoted by $r_1,\ldots,r_n$, where $g$ is the inverse link function from this model. The idea is that the residuals contain information about the treatment differences.
From these residuals follow the residual-based test population-based MCP-Mod type statistics
\begin{eqnarray}\label{eq:MCP_resid}
  S_2(\mathbf{Z}) =\underset{m\in\{1,...,M\}}{\max} T_{2m}(\mathbf{Z}), \; \mathrm{where} \; T_{2m}(\mathbf{Z}) = \frac{\mathbf{c}_m'\bar{\boldsymbol{r}}_{\mathbf{Z}}}{\sqrt{\sum_{j=0}^{k-1}c_{m,j}^2 s_j^2/n_j}},
\end{eqnarray}
where $\bar{\boldsymbol{r}}_{\mathbf{Z}}$ is the vector of residual group means $\bar{\boldsymbol{r}}_{\mathbf{Z}}:=(\bar{r}_0,...,\bar{r}_{k-1})$ and respective empirical variances $\boldsymbol{s}^{\mathbf{2}}_{\mathbf{Z}}=(s^2_0,...,s^2_{k-1})$.
The sample sizes per group, as weight vector, and candidate models are used to calculate the contrasts $\mathbf{c}_1,...\mathbf{c}_M$ \citep{pinh:born:glim:2014}.

\textbf{Monte Carlo Randomization Test} \\
Both Test Statistics $S_{1}$ and $S_{2}$ can be employed in a Monte Carlo randomization test. 
To construct a Monte Carlo randomization test using either of the two test statistics \eqref{eq:MCP} or \eqref{eq:MCP_resid} described above. We repetitively sample a treatment assignment sequence $\mathbf{Z}_l = (Z_{l,1}, \ldots, Z_{l,n})$ for $l=1, \ldots, n_{\text{rand}}$ from a reference set $\mathcal{R}_Z$ using the randomization probability distribution \(\mathcal{P}(\mathbf{Z})\) used in the trial, where $n_{\text{rand}}$ denotes the number of re-randomizations in the randomization test. The reference set $\mathcal{R}_Z$ contains all potential treatment assignment sequences that could have occurred under the randomization procedure employed, see more in Section \ref{Sec:RandProcedures}.

The hypothesis \eqref{eqn:h0} implies that the observed responses $y_{i, z_{i}}$ are independent of the treatment assignment $z_{i}$. Therefore, for each randomization, the vector of observed treatment responses $\mathbf{Y} = (y_1, \ldots, y_n)$ remains unchanged, while the vector of treatment assignments \(\mathbf{Z}_l \) alters. Using the original responses $\mathbf{Y}$, we compute the test statistic \eqref{eq:MCP} or \eqref{eq:MCP_resid} based on $\mathbf{Z}_l$ and yield values $S_{i}(\mathbf{Z}_1), \ldots, S_{i}(\mathbf{Z}_{n_{\text{rand}}})$ for $i=1,2$ and $l=1, \ldots, n_{\text{rand}}$.
With these test statistics, we can compute the one-sided p-value as
\begin{equation}
    \hat{p} = \frac{\sum_{l=1}^{n_{\text{rand}}} \textbf{1}(S_{i}(\mathbf{Z}_l) \geq S_{\text{i}}(\mathbf{Z}_{obs}))}{n_{\text{rand}}}.
\end{equation}

The different estimation techniques from Section \ref{Sec:Estimation} and test statistics from this section lead to five different MCP-Mod test statistics that are explored in this paper:
\begin{enumerate}
    \item Population-based test, as introduced in Section \ref{Sec:MCP-Mod} (Population-based Test),
    \item randomization test based on the generalized MCP-Mod test statistic which uses MLE (Randomisation Test using MLE),
    \item randomization test based on the residual approach which uses MLE  (Randomisation Test using MLE (residual-based))
    \item randomization test based on the generalized MCP-Mod test statistic which uses penalised MLE (Randomisation Test using penalized MLE) and 
    \item randomization test based on the residual approach which uses penalised MLE (Randomisation Test using penalized MLE (residual-based)). 
\end{enumerate}
From now on, the numbering and names in brackets refer to the five different tests in this paper.
In the next section, we explore different randomization procedures, as this can have an impact on the performance of the randomization test.


\subsubsection{Randomization Procedures}\label{Sec:RandProcedures} 

We now introduce three different randomization procedures and explain how they influence the study design and analysis. The procedures vary by their degree of restriction. For a more general overview of different randomization procedures, see Rosenberger \& Lachin (2016)\cite{rosenberger2016randomization} and Berger et al. (2021) \cite{berger2021roadmap}. 

The first procedure is \textit{complete randomization (CR)}, or unrestricted randomization. In the scenario of a two-armed experiment with a binary endpoint, CR can be seen as a simple coin toss. In the multi-armed context, we use a $k$-sided die that resembles drawing from a multinomial distribution for each patient $i$.  
The main advantage of CR lies in its ability to prevent selection bias because each patient's assignment is independent of previous assignments. 
In CR, the total number of possible randomization
sequences is given by $|\mathcal{R}_Z| = k^n$ including those that allocate all participants to a single treatment group.
As a result, in CR, every treatment option has an equal probability of $1/k$ for each patient. This ensures that there is no reason to believe that the next patient will be assigned to any particular treatment with a higher probability than any other, a feature not guaranteed by other randomization methods. 
However, a significant drawback of CR is the lack of control over the final allocation, which can result in deviations from the targeted distribution. Although each treatment arm should, in theory, receive roughly $n/k$ patients, in practice there can be notable discrepancies, especially in smaller samples. In extreme cases, an arm might end up with very few or even no patients, which is particularly problematic in randomized controlled trials. Consequently, Schulz et al. (2002) \cite{schulz2002unequal} recommend employing CR only for trials with moderate to large sample sizes, e.g., $n > 200$.

The other two procedures we consider in this paper address this issue by enforcing a pre-specified final allocation. One such procedure, the \textit{random allocation rule (RA)}, can be formalized as an urn design. In this design, the number of balls for each treatment corresponds to the number of patients intended for that treatment. We then would repeatedly draw balls at random, without replacement until the urn is empty, ensuring each treatment group receives its target quota of patients. The number of different allocation sequences for the RA procedure is $|\mathcal{R}_Z|=\binom{n}{n_0\;n_1\;\cdots\;n_{k-1}} = \frac{n!}{n_0!n_1!\cdots n_{k-1}!}$, where $n_j\ge 1$ is the final target sample size for the $j$-th group, $\sum_{j=0}^{k}n_j=n$, and each sequence in the reference set is equally likely to occur, with probability $\binom{n}{n_0\;n_1\;\cdots\;n_{k-1}}^{-1}$. While the RA procedure achieves exactly the target allocation numbers for the treatment groups, it may still result in non-negligible imbalances between treatments during the trial, which can pose challenges when adjusting for time trends. 
However, this procedure has several disadvantages. For one, it can lead to selection bias, as the allocation of patient $i$ is not independent of previous allocations. In extreme cases, when the remaining balls in the urn correspond to only one treatment, this can result in guaranteed predictions. 

The third procedure, the \textit{permuted block design (PBD)}, addresses this issue by repeatedly executing the random allocation rule with smaller blocks or urns. This procedure involves selecting a block size $B$ such that the total number of patients $n$ can be divided into $n/B$ identical blocks. These smaller blocks mirror the larger urn used in RA in terms of the proportion of patients allocated to each treatment. By carefully choosing a sufficiently small block size, this method helps prevent substantial imbalances throughout the trial. Due to this advantage, PBD is the most used design in practice \citep{sverdlov2023randomization}. However, it also increases the predictability of the allocation of the next patient towards the end of each block, thereby raising the risk of selection bias when the block size is too small.
The number of different randomization sequences for a PBD with $\ell$ blocks of length $m$ (such that $\ell m=n$ and each block has the allocation ratio $m_0:m_1:\cdots:m_k$, with $\sum_{j=0}^km_j=m$) is  $\binom{m}{m_0\;m_1\;\cdots\;m_{k-1}}^\ell$ and all sequences are equally likely.

To illustrate the three randomization procedures, considered the clinical trial example from Section~\ref{Sec:Trial} with $n=49$. We can have 7 blocks of length 7, each with $1:2:2:2$ allocation ratio. The total number of randomization sequences for PBD is $\binom{7}{1\;2\;2\;2}^7=630^7\approx 3.94 \cdot 10^{19}$, then for RA is $\binom{49}{7\;14\;14\;14} = \frac{49!}{7!14!14! 14!} \approx 1.82 \cdot 10^{26}$ and for CR it is $7^{49}\approx 2.57 \cdot 10^{41}$.  
Each randomization procedure results in a distinct probability distribution $\mathcal{P}$. The set of sequences $\mathcal{R}_Z$ with non-zero probabilities varies for each procedure, as does the probability assigned to each sequence. Specifically, the set of possible sequences for PBD is a subset of those for RA, which in turn is a subset of the sequences for CR. Consequently, when conducting a randomization test, sequences are drawn from different reference sets, each with its unique probability distribution based on the randomization procedure used to generate the sample. The impact of this will be explored in detail through our simulation studies in Section \ref{Sec:Simulations}.

\section{Simulations}\label{Sec:Simulations}

\subsection{Simulations for Binary Endpoint}\label{Sec:MainSimulation}

In this section, we use the clinical trial example from Section \ref{Sec:Trial} and consider in total 14 scenarios to investigate the properties of randomization-based tests for MCP-Mod, see Table \ref{tab:DataScenarios}. 
We investigate three sample size scenarios. Doubling the original trial size provides insights into a moderate sample, while increasing the trial size tenfold allows us to examine asymptotic properties. This approach is driven by the notably small sample size of the original trial, which is why we do not consider exploring smaller trial sizes. 
For comparability across scenarios the response probability at the highest dose $p_{k-1}$ was changed for each sample size to achieve the same (asymptotic) power, for population based MCP-Mod, as for the scenario with the smallest trial size $(n=49)$, which used $p_{k-1}=0.8$. Calculations were performed using the MCP-Mod design application \citep{sun2025mcpmod}, resulting in success probabilities of \( p_{k-1}=0.61 \) for \( n=98 \) and \( p_{k-1}=0.364 \) for \( n=490 \). Differences in the reported power results in Table \ref{tab:AllResults} arise due to complete separation issues and the distinction between finite-sample and asymptotic properties. 
In addition to PBD, we examine the performance of RA for all and CR for the largest trial sizes only. For the two smaller trial sizes, the likelihood of allocating very few patients to the control arm is too high when using CR. In the case of the smallest sample size, there is even a possibility that only one or none of the patients would be allocated to the control arm. 
Lastly, we also rerun all the scenarios with the inclusion of a time trend. 

\begin{table}[htb!]
    \centering
    \begin{tabular}{ccc}
        Sample Size & Randomization Procedure & Time Trend \\
        \hline
        \textbf{49} & \{RA, \textbf{PBD}\} & \{\textbf{No}, Yes\} \\
        98 & \{RA, PBD\} & \{No, Yes\}\\
        490 & \{RA, PBD, CR\} & \{No, Yes\} \\
    \end{tabular}
    \caption{The 14 data generating scenarios of the first simulation study. The original scenario described in Section \ref{Sec:Trial} is depicted in bold.}
    \label{tab:DataScenarios}
\end{table}

\begin{figure}[hbt!]
    \centering
    \includegraphics[width=0.7\linewidth]{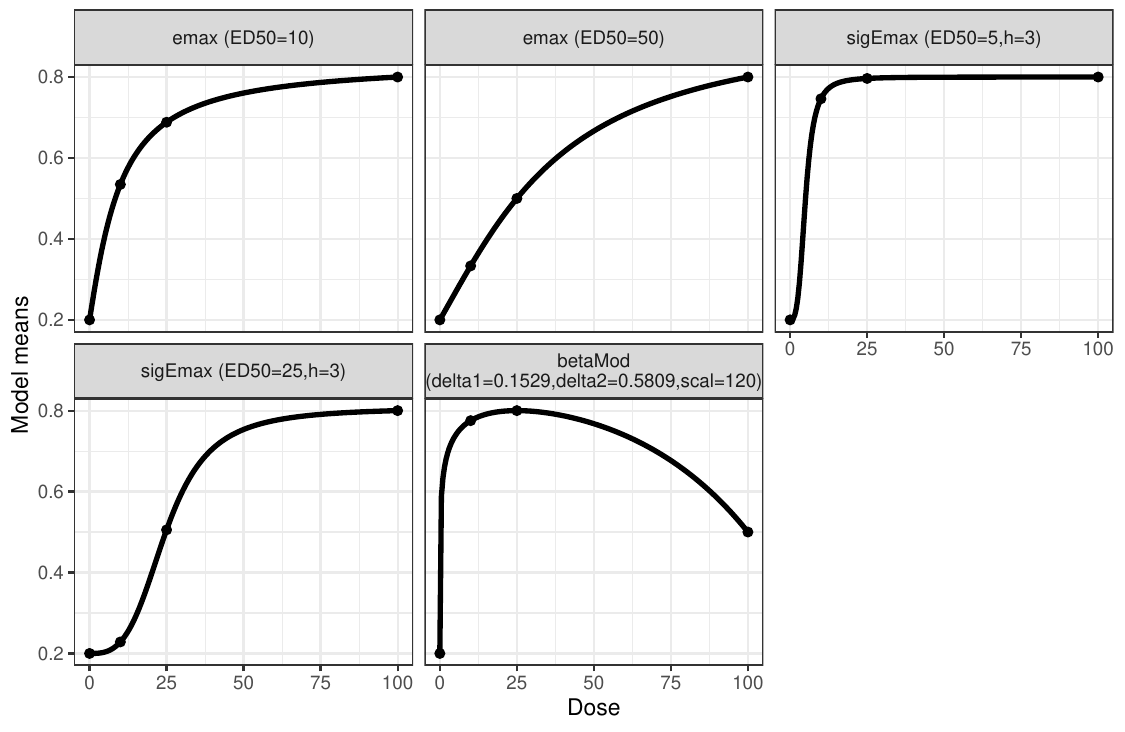}
    \caption{Five candidate models for MCP-Mod (transformed from a logit scale to a probability scale for easier interpretability).}
    \label{fig:CandModels}
\end{figure}

For the analysis we use the following candidate dose-response models: two Emax models $f(x, \theta_0, \theta_1, \theta_2) = \theta_0 + \theta_1 x / (\theta_2 + x)$, two Sigmoid Emax models $f(x, \theta_0, \theta_1, \theta_2, h) =  \theta_0 + \theta_1 x^{h} / (\theta_2^{h} + x^{h}) $ and a beta model $f(x, \theta_0, \theta_1, \delta_1, \delta_2) =  \theta_0 + \theta_1 B(\delta_1, \delta_2) (x/120)^{\delta_1} (1-x/120)^{\delta_2}$, where $B()$ is the Beta-function. For how the specific parameters $\theta_0, \theta_1, \theta_2, h, \delta_1 \text{ and } \delta_2$ are chosen we refer to Bretz et al. (2005) \cite{bret:pinh:bran:2005} and  Pinherio et al. (2014) \cite{pinh:born:glim:2014}. In this paper we choose them to represent a range of possible different dose-response relationships, see Figure \ref{fig:CandModels}.We even allow for a possibly non-monotone dose-response relationship. 

We choose the Emax model with $\theta_2=10$ as the ground truth for the 14 simulation scenarios to provide a common basis for comparing the population-based generalized MCP-Mod approach with our proposed randomization-based inference extensions. Since all methods utilize the same candidate models, the observed differences in power should not be driven by the choice of the underlying truth but rather by the differences between the testing approaches.
For the Emax model with $\theta_2=10$, we obtain $\theta_0 = \log(p_{0}/(1-p_{0}))$ and determine $\theta_1$ by solving
\begin{equation}
    \log\frac{p_{k-1}}{1-p_{k-1}}  =  \log\frac{p_{0}}{1-p_{0}} +  \theta_1 \cdot \frac{d_{k-1}}{10+d_{k-1}},
\end{equation}
e.g. $\theta_0 \approx -1.39$, $\theta_1 \approx 3.05$  and $\theta_2 = 10$ for the clinical trial example with $n=49$ and $p_{k-1}=0.8$.
We then plug the vector of treatment sequences $Z=(Z_1,...,Z_n)$ into the Emax function in order to derive individual success probabilities. We also examine the influence of one covariate. Multiple covariates can be utilized to incorporate patient characteristics, including demographic factors such as age, gender, and race. Baseline measurements pertaining to the disease in question, such as weight, blood pressure or race can also be included \citep{ye2022toward}. In addition biomarkers can be added as important covariates \citep{sun2022comparing}. In all scenarios of this simulation study we include just one covariate. For this, we draw values $x_{1},...x_{n}$ from a standard normal distribution and subsequently calculate the values $y_i=f(Z_i,\theta)+0.6 x_{i}$ for $i=1,..,n$. Applying the inverse of the logit function to these values provides the individual probabilities of success $\gamma_1,...,\gamma_n$.
The factor of $0.6$ is selected so that the covariate has a predictive strength of $AUC=0.66$ for the control group, which is motivated by a real trial example \cite{sun2022comparing}. 

To investigate the influence of a linear time trend $t_i$, we add a time-varying value to the original probability vector $\gamma$, where $\gamma$ is the vector of individual success probabilities for each patient $i = 1, \dots, n$. To ensure all probabilities fall within the interval $(0,1)$, we adjust the probability as follows:
\[
\gamma_i^t = \max(0,\min(\gamma_i + t_i, 1))
\]
This ensures that the time-adjusted probabilities $\gamma_t$ remain valid, without exceeding the upper bound of 1. We define the time trend as $t_i = 0.4 \cdot i/n - 0.2$ for $i=1,..,n$.
Across all those 14 different data generating scenarios we compare the five different tests from Section \ref{Sec:GenervalvsResidu}. For the randomization tests $n_{rand}=1,000$ re-randomizations are performed and each scenario is simulated $n_{sim}=10,000$ times.

\begin{figure}[htb!]
    \centering
    \includegraphics[width=0.9\linewidth]{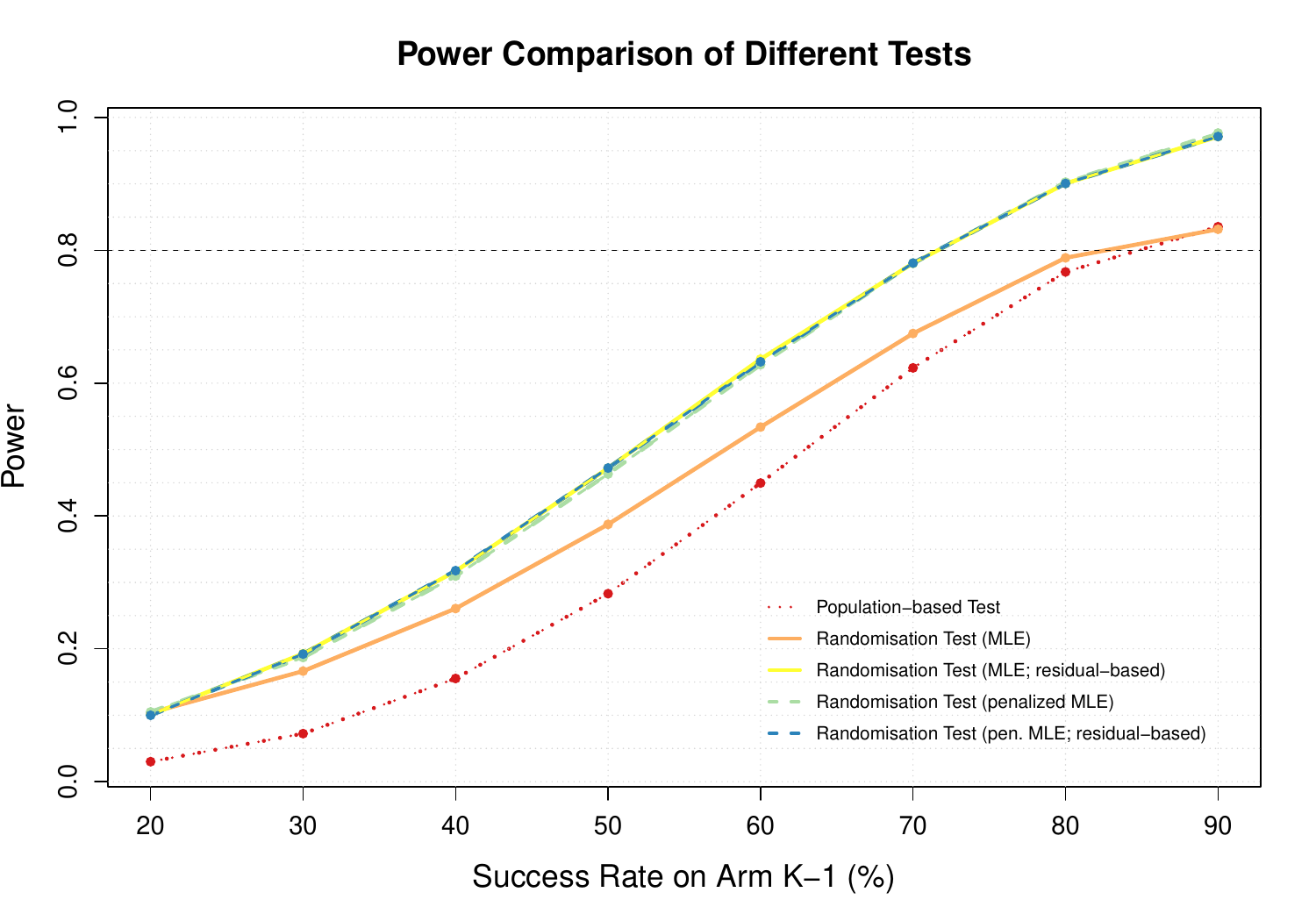}
    \caption{Power of the five methods in the original scenario, where $n=49$, $p_0=0.2$, $p_{k-1}={0.2,0.3,...,0.9}$ and PBD is used as randomization procedure.}
    \label{fig:OrgPower}
\end{figure}

We begin by comparing the results for the original scenario described in Section \ref{Sec:Trial}. In Figure \ref{fig:OrgPower}, we assume a success rate of \( p_0 = 0.2 \) for the control arm and varied the success rate for the highest treatment arm, \( p_{k-1} \in \{0.2, 0.3, \ldots, 0.9\} \). The x-axis represents the success rate, while the y-axis shows the power level across the five different methods. Power is defined as the proportion of simulation runs where the p-value is less than the significance level of 0.1:
\[
\text{Power} = \frac{\#(\text{p-value} < 0.1)}{\text{total simulations under the alternative hypothesis}}.
\]
Similarly, the type-I error rate is defined as the proportion of simulations under the null hypothesis where the p-value is less than the siginificance level 0.1:
\[
\text{Type-I Error Rate} = \frac{\#(\text{p-value} < 0.1)}{\text{total simulations under the null hypothesis}}.
\]
In simulations involving complete or quasi-complete separation, methods using the MLE still provide a p-value, especially the population-based test. However, it is crucial to recognize that the p-values of these methods are not meaningful due to the software's limitations in addressing complete separation; the standard implementation for glm in R will always return ML estimates, even if they don't exist. Usually estimates in this situation will be large with large standard errors. Nevertheless, we include these p-values in the calculation of both power and type-I error rates to ensure consistency across all five methods and to accurately reflect the method's behavior in practical applications.

\begin{figure}
    \centering
    \includegraphics[width=0.8\linewidth]{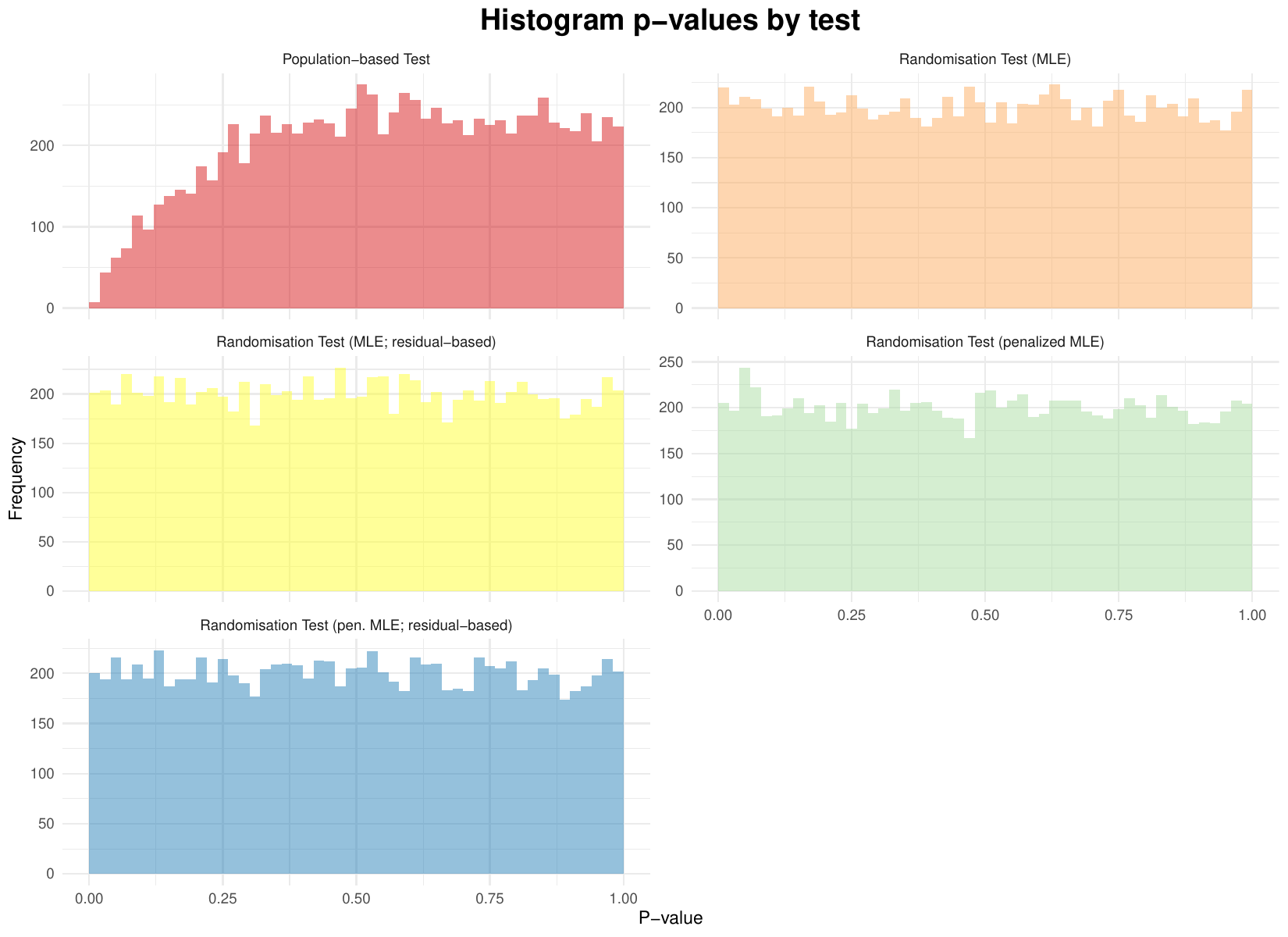}
    \caption{Empirical distribution of the p-values under the null-hypothesis of the original scenario, i.e. where $n=49$, $p_0=p_{k-1}=0.2$ and PBD is used as randomization procedure for all five tests.}
    \label{fig:pvalsplot}
\end{figure}

Figure \ref{fig:OrgPower} demonstrates that the two randomization tests using the residual-based approach achieve similar power levels, along with the randomization test using the MCP-Mod test statistic \eqref{eq:MCP} but employing penalized MLE rather than the MLE. All three methods consistently exhibit higher power across all success rates compared to the population-based test approach.
Using the population-based tests or the randomization-based test with the MCP-Mod statistic and maximum likelihood estimation results in a loss of power due to complete separation scenarios. More importantly, in the case of complete separation in an actual trial (with an 18\% chance of occurrence, as noted in Section \ref{Sec:Estimation}), these methods would not yield meaningful results.  
In addition, all randomization tests exhaust the significance level, as depicted in Figure \ref{fig:pvalsplot} and detailed in Table \ref{tab:AllResults} even when maximum likelihood estimation is used (i.e., when complete separation may occur). In contrast, the population-based test fails to do so.

Lastly, to comprehensively evaluate the computational efficiency of the various statistical tests under consideration, all analyses were executed on a high-performance computational server. This server is equipped with an Intel(R) Xeon(R) Gold 6254 CPU, with a base clock speed of 3.10 GHz and featuring 18 physical cores and 36 logical processors enabled through hyper-threading. The analyses were conducted using \texttt{R} version \texttt{4.3.1}.
Randomization tests, owing to the iterative calculation of the test statistic, are often computationally intensive. A key objective of adopting a residual-based approach is to mitigate this computational burden, particularly for the randomization test that relies on the MCP-Mod test statistic. The residual-based approach circumvents the necessity of refitting the logistic regression model during each re-randomization iteration. Furthermore, the MCP-Mod-based randomization test demands a correction using penalized MLE estimation. This is inherently more demanding computationally compared to standard MLE, but for the residual-based approach it only needs to be calculated once. 
On our computational setup, the randomization test using the MCP-Mod test statistic with penalized MLE estimations requires an average of 7 seconds to complete (averaged over 10,000 simulations). In contrast, the two residual-based approaches run in less than 0.5 seconds each, an speedup of over 14 times.
Given the substantial reduction in the runtime, coupled with the advantageous properties previously detailed, the residual-based approach is the preferred method for conducting randomization tests.

The analysis of the additional 13 scenarios, as presented in Table \ref{tab:AllResults}, corroborates the earlier observations and provides further insights.
Firstly, regarding the impact of time trends, it is noted that power is not negatively affected by the presence of a time trend. If anything, there is a slight increase in power in scenarios where time trends are present. However, the type-I error rate, in scenarios where PBD is used as randomization procedure, tends to be deflated. Notably, for the largest sample size, the type-I error rate of the population-based test remains deflated under PBD but not under RA and CR when a time trend exists. This observation aligns with findings from previous studies \cite{rosenkranz2011impact, berger2021roadmap}. The randomization tests control and exhaust the type-I error for all randomization procedures unlike the population-based test.

Secondly, when comparing the original scenario, where PBD is the randomization method, to other scenarios with the smallest sample size of $n=49$, all randomization tests exhibit higher power when PBD is used as the randomization procedure compared to RA. This increased power with PBD is also observed in larger sample sizes of $n=98$ and $n=490$. CR shows even slightly lower power compared to RA.
Lastly, for larger sample sizes, we see that randomization tests asymptotically control type-I error extremely effectively. The power difference between the population-based MCP-Mod test statistic and residual-based tests using the same estimator becomes negligible for the largest sample size of $n=490$. This finding is reassuring for the residual-based approach.

In conclusion, our evaluation highlighted that residual-based randomization tests offer superior power and robust type-I error rate control compared to the population-based MCP-Mod method. These tests effectively mitigate issues of complete separation when penalized MLE are used, allow the inclusion of multiple covariates and are computationally efficient, making them well-suited for clinical trials with varied conditions and sample sizes. 
Additionally, while the presence of time trends and different randomization methods can affect performance, residual-based tests consistently demonstrated reliability and effectiveness across various scenarios.
Although we also examined a population-based MCP-Mod approach using penalized MLEs during our investigation, our primary focus in this paper remained on the randomization-based framework. While the population-based MCP-Mod approach using penalized MLEs helps mitigate complete separation and offers some power gains, it remains less effective than randomization-based methods. Notably, the randomization-based inference framework ensures type-I error rate control, even in the presence of time trends, and generally exhibits greater robustness across various scenarios.

\begin{table*}[htbp]
\small
\renewcommand{\arraystretch}{0.875}
\def\d{\hphantom{0}}
\caption{Reporting type-I error rate (success rate of $p_0=...=p_{k-1}=20$\% for all doses) and power (success rate of $p_{k-1}=80$\% when $n=49$, $p_{k-1}=61$\% when $n=98$ and $p_{k-1}=36.1$\% when $n=490$ on the highest dose) for all scenarios and 
tests: 1 = Population-based Test, 2 = Randomisation Test using MLE, 3 = Randomisation Test using MLE (residual-based), 4 = Randomisation Test using
penalized MLE, 5 = Randomisation Test using penalized MLE
(residual-based). The Monte-Carlo error for power is below 0.45\% and type-I error even below 0.31\% \citep{morris2019using}. \label{tab:AllResults}}
\begin{tabular*}{\textwidth}{@{\extracolsep{\fill}}llllcc@{}}
\toprule
Sample Size & Randomization Procedure & Time Trend & Test & Type-I Error & Power \\
\midrule
49 & RA & No & 1 & 2.72 & 76.85 \\
49 & RA & No & 2 & 9.75 & 73.74 \\
49 & RA & No & 3 &  9.85 & 86.65 \\
49 & RA & No & 4 & 9.90 & 86.73 \\
49 & RA & No & 5 & 10.00 & 86.64 \\ \hdashline 
49 & RA & Yes & 1 & 2.49 & 76.40 \\
49 & RA & Yes & 2 & 9.90  & 75.31 \\
49 & RA & Yes & 3 & 9.83 & 87.49 \\
49 & RA & Yes & 4 & 9.52 & 87.84 \\
49 & RA & Yes & 5 & 9.70 & 87.52 \\ \hdashline

49 & PBD & No & 1 & 3.03 & 76.75 \\
49 & PBD & No & 2 & 10.37 & 78.87 \\
49 & PBD & No & 3 & 10.06 & 90.07 \\
49 & PBD & No & 4 & 10.50 & 90.26 \\
49 & PBD & No & 5 & 9.99 & 90.06 \\ \hdashline
49 & PBD & Yes & 1 & 1.91 & 76.89 \\
49 & PBD & Yes & 2 & 9.52 & 80.97 \\
49 & PBD & Yes & 3 & 10.28 & 90.81 \\
49 & PBD & Yes & 4 & 9.85 & 91.28 \\
49 & PBD & Yes & 5 & 10.21 & 90.58 \\ \hdashline

98 & RA & No & 1 & 5.98 & 84.55 \\
98 & RA & No & 2 & 10.39 & 81.37 \\
98 & RA & No & 3 & 10.45 & 83.73 \\
98 & RA & No & 4 & 10.36 & 83.69 \\
98 & RA & No & 5 & 10.43 & 83.77 \\ \hdashline
98 & RA & Yes & 1 & 5.49 & 84.72 \\
98 & RA & Yes & 2 & 9.95 & 83.53 \\
98 & RA & Yes & 3 & 9.93 & 85.20 \\
98 & RA & Yes & 4 & 10.02 & 85.30 \\
98 & RA & Yes & 5 & 9.85 & 83.28 \\ \hdashline

98 & PBD & No & 1 & 5.88 & 85.02 \\
98 & PBD & No & 2 & 10.33 & 86.01 \\
98 & PBD & No & 3 & 10.10 & 87.89 \\
98 & PBD & No & 4 & 10.51 & 87.75 \\
98 & PBD & No & 5 & 10.06 & 87.77 \\ \hdashline
98 & PBD & Yes & 1 & 4.96  & 85.28 \\
98 & PBD & Yes & 2 & 9.98 & 87.89 \\
98 & PBD & Yes & 3 & 10.28 &  89.15 \\
98 & PBD & Yes & 4 & 10.02 & 89.31 \\
98 & PBD & Yes & 5 & 10.21 & 89.27 \\ \hdashline

490 & RA & No & 1 & 9.49 & 85.09 \\
490 & RA & No & 2 & 9.96 & 81.44 \\
490 & RA & No & 3 & 10.00 & 81.84 \\
490 & RA & No & 4 & 10.01 & 81.44 \\
490 & RA & No & 5 & 9.97 & 81.75 \\ \hdashline
490 & RA & Yes & 1 & 9.53 & 83.51 \\
490 & RA & Yes & 2 & 10.08 & 81.37 \\
490 & RA & Yes & 3 & 9.99 & 81.42 \\
490 & RA & Yes & 4 & 9.96 & 81.44 \\
490 & RA & Yes & 5 & 10.19 & 81.71 \\ \hdashline

490 & PBD & No & 1 & 9.32 & 85.03 \\
490 & PBD & No & 2 & 9.81 & 85.47 \\
490 & PBD & No & 3 & 9.78 & 85.65 \\
490 & PBD & No & 4 & 9.84 & 85.53 \\
490 & PBD & No & 5 & 9.90 & 85.70 \\ \hdashline
490 & PBD & Yes & 1 & 7.66 & 84.35 \\
490 & PBD & Yes & 2 & 9.56 & 86.29 \\
490 & PBD & Yes & 3 & 9.56 & 86.45 \\
490 & PBD & Yes & 4 & 9.50 & 86.25 \\
490 & PBD & Yes & 5 & 9.69 & 84.40 \\ \hdashline

490 & CR & No & 1 & 9.29 & 84.65 \\
490 & CR & No & 2 & 9.76 & 80.73 \\
490 & CR & No & 3 & 9.93 & 81.16 \\
490 & CR & No & 4 & 9.81 & 80.70 \\
490 & CR & No & 5 & 9.79 & 81.27 \\ \hdashline
490 & CR & Yes & 1 & 9.05 & 83.53 \\
490 & CR & Yes & 2 & 9.90 & 81.41 \\
490 & CR & Yes & 3 & 9.93 & 81.45 \\
490 & CR & Yes & 4 & 9.80 & 81.24 \\
490 & CR & Yes & 5 & 10.10 & 81.71 \\ 
\bottomrule
\end{tabular*}
\end{table*}

\subsection{Pharmacometric Simulation Study}\label{Sec:Pharmaco}

In this section, we explore the potential of randomization tests for MCP-Mod in a simulation scenario motivated by a pharmacometric model with a continuous endpoint. We conduct a simulation study inspired by the example presented by Buatois et al. (2021) \cite{buatois2021clrt}, who proposed cLRT-Mod, a pharmacometric adaptation of the MCP-Mod methodology. The simulation in this section does not use a population model to simulate new outcome data for each simulated trial, but rather generates all potential outcomes for every patient once based on a pharmacometric model using a quasi-Monte Carlo method. Then each simulated trial only differs by the generated treatment sequence, i.e which potential outcome is observed for each patient. To our knowledge, the synergies between pharmacometric modeling and randomization tests have not yet been explored. Pharmacometric modeling often focuses on modeling the individual dose-response curves (via use of covariates and random effects), this makes it ideally suited to generate potential outcomes under all doses for an individual patient.

We base our simulation study on the hypothetical phase II dose-finding trial for the development of a new monoclonal antibody for the treatment of wet-AMD \citep{rosenfeld2006ranibizumab} presented by Buatois et al. (2021) \cite{buatois2021clrt}. This disorder is characterized by the growth of abnormal blood vessels beneath the retina, damaging the macula and impairing visual acuity (VA). The aim of the new treatment is to target vascular endothelial growth factor (VEGF) in the retina to improve VA. The VA improvement is measured using the early treatment diabetic retinopathy study (ETDRS) chart, with scores based on the number of readable letters out of 70. The trial duration is set to 12 months with five parallel arms (placebo, or one of the following doses: 100, 200, 400, and 1000 $\mu$g). Observations are collected at baseline, day 7, and monthly for 12 months. In our simulation study, the baseline observation is included as a covariate, and the month 12 observation as primary endpoint. All other monthly observations are not considered in the analysis.

\begin{figure}[htb!]
    \centering
    \includegraphics[width=0.7\linewidth]{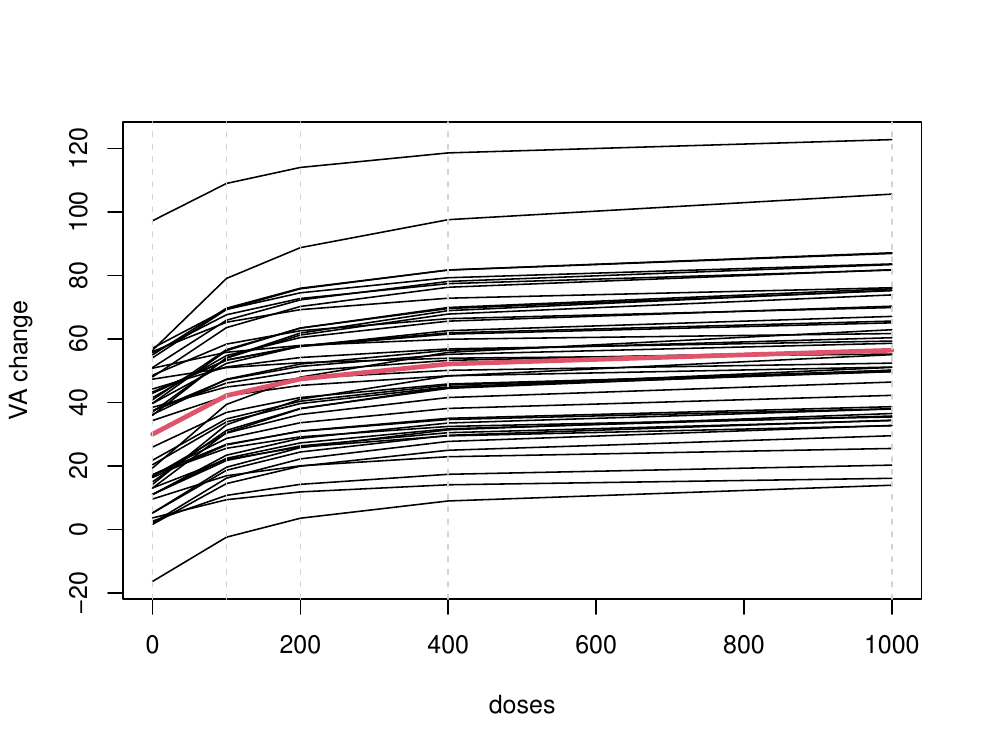}
    \caption{Individual dose-response curves for the 50 patients. The average of the individual curves is shown in red and the grey lines represent the dose levels that are included in the trial.}
    \label{fig:IndividualDR}
\end{figure}

For our first data-generating scenario, we use the case of a few patients (\(n = 50\)) and a strong drug effect \cite{buatois2021clrt}. The set of candidate models includes flat, linear, log-linear, Emax, and Sigmoid functions. The parameters for these models can be found in Buatois et al. (2021) \cite{buatois2021clrt}. 
We generated all potential outcomes under the Emax model with $\theta_2=150$ and the individual dose-response curves and thus potential outcomes for each patient generated from the pharmacometric model are shown in Figure \ref{fig:IndividualDR},  the average dose-response curve is shown in red.
Generation of these potential outcomes $y_{i,d_k}$ using the pharmacometric model for all patients and dose levels, with $i=1,\ldots,50$ and $k=0,\ldots,4$, was done using a quasi-random sample from the model using a Richtmyer sequence \citep{richtmyer1951evaluation}. The resulting individual dose-response curves shown in Figure \ref{fig:IndividualDR}. 

Given the  dataset of potential outcomes with dimensions $50 \times 5$, we conduct clinical trial simulations applying two randomization procedures: RA with an equal allocation of 10 patients per treatment, and PBD with a block size of 10, allocating 2 patients per treatment across 5 blocks. On these clinical trial realizations we perform statistical analyses using both the population-based test and the randomization-based approach, evaluating the results both with and without baseline measurements as covariates, see Table \ref{tab:pharmaco}.
\begin{table}[htbp]
    \centering
    \begin{tabular}{lllll}
         Scenario & Randomization Procedure & Covariates incl. &  Test  & Power (\%) \\ \hline
         1 & RA & No & Population &  88.00 \\
         1 & RA & No & Randomization  &  87.80 \\
         1 & RA & Yes & Population & 98.60 \\
         1 & RA & Yes & Randomization & 98.57 \\ \hdashline
         1 & PBD & No & Population & 87.44 \\
         1 & PBD & No & Randomization & 85.58 \\
         1 & PBD & Yes & Population & 98.32 \\
         1 & PBD & Yes & Randomization & 97.83 \\ \hdashline
         2 & RA & No & Population & 88.59 \\
         2 & RA & No & Randomization & 88.59 \\
         2 & RA & Yes & Population & 98.66 \\
         2 & RA & Yes & Randomization &  98.62 \\ \hdashline
         2 & PBD & No & Population & 91.96 \\
         2 & PBD & No & Randomization &  97.67 \\
         2 & PBD & Yes & Population &  98.89 \\
         2 & PBD & Yes & Randomization &  98.93 \\
         &&&&\\
        Scenario & Randomization Procedure & Covariates incl. &  Test  & Type-I Error (\%) \\ \hline
         3 & RA & No & Population & 4.88 \\
         3 & RA & No & Randomization & 4.87  \\
         3 & RA & Yes & Population &  5.15 \\
         3 & RA & Yes & Randomization & 5.19  \\ \hdashline
         3 & PBD & No & Population &  5.62 \\
         3 & PBD & No & Randomization &  4.86  \\
         3 & PBD & Yes & Population & 5.36 \\
         3 & PBD & Yes & Randomization &  4.69 \\
    \end{tabular}
    \caption{For the randomization tests $n_{rand}=1,000$ re-randomizations are performed and the scenario was simulated $n_{sim}=10,000$ times. The Monte-Carlo standard error for power and type-I error is below 0.4\% \citep{morris2019using}.}
    \label{tab:pharmaco}
\end{table}
Our findings indicate that the randomization-based test exhibits similar power to the population-based MCP-Mod test, regardless of the randomization procedure employed (RA or PBD) and irrespective of whether covariates are included. However, incorporating baseline measurements as covariates consistently enhances the power of both tests by approximately 10\%.

In the second scenario presented in Table \ref{tab:pharmaco}, we sort the potential outcomes based on the baseline observation, which can be interpreted as a positive time trend. One potential interpretation is that patients who join the trial later tend to have higher baseline observations, and are in better health. This could be because patients whose visual acuity is most severely impacted by wet-AMD might enroll in the trial sooner. These patients might either volunteer earlier due to the severity of their condition or be prioritized for selection in hospitals, thus causing those with better visual acuity to enroll later.

The power of the randomization-based test and population based test is basically identical when RA is used as the randomization procedure, see Table \ref{tab:pharmaco}. 
However, the results demonstrate that the randomization test controls for the time trend (even if baseline value is not included as covariate) when the PBD is used as a randomization procedure and the population-based test does not. The difference in power between the two types of tests is over 5\%.

In the third and last scenario, we aim to investigate the type-I error rate. For this, we set:
$y_{i,d_{0}} = y_{i,d_{1}} = \ldots = y_{i,d_{4}}$
for all \(i=1,\ldots,50\). Hence, the outcome of each individual patient is independent of their treatment assignment. All differences between the set of all potential outcomes are now due to baseline differences between patients or random effects. Table \ref{tab:pharmaco} shows that the randomization test controls type-I error better. The population-based test exhibits inflation of the type-I error when PBD was used as the randomization procedure. This is in line with our findings from the simulation study in Section \ref{Sec:MainSimulation}.

In conclusion, motivated by a pharmacometric model, this simulation study
confirmed the previous finding that randomization tests control type-I error and, when combined with PBD, account for time trends.

\section{Discussion}\label{Sec:Discussion}

In this paper, we explored randomization-based inference methods as alternatives to the conventional population-based test for MCP-Mod analysis in clinical trial settings, with a focus on addressing challenges related to small sample sizes, complete separation and time trends. Below, we summarize the key findings of our studies and suggest directions for future research.

\subsection{Key Findings}

One of the primary challenges we encounter in the population-based MCP-Mod is the issue of complete separation in logistic regression, particularly in studies with small sample sizes. This problem often results in cases where the MLE either does not exist, see Section \ref{Sec:Estimation}. We have demonstrated that using penalized MLE with randomization-inference addresses these issues effectively, ensuring finite and meaningful estimates. This improvement was shown to be especially significant in small to medium-sized samples, where the likelihood of encountering complete separation is higher, see Section \ref{Sec:MainSimulation}.

Our simulation studies further revealed that randomization-based tests can exhibited higher power compared to the population-based MCP-Mod, particularly in trials with small to medium sample sizes. The power increase was especially notable in small samples, where the randomization test showed up to a 14\% improvement over the population-based test, as demonstrated in Section \ref{Sec:MainSimulation}. This enhanced power was consistent across various data-generating scenarios, including those with and without covariates, as detailed in Section \ref{Sec:Pharmaco}.

Moreover, the randomization tests controlled type-I error across all tested randomization procedures: complete randomization, random allocation rule, and permuted block randomization. This accuracy was consistent across different sample sizes and data-generating scenarios, even in the presence of time trends. In contrast, the population-based test exhibited deflated type-I error rates under certain conditions, particularly when PBD was used in the presence of time trends, see Section \ref{Sec:MainSimulation}.

In this paper, we compared four randomization tests based on MCP-Mod to the population-based MCP-Mod, leading to the following conclusions: The residual-based randomization tests based on test statistic $S_2$ performed as well as, or better than, those based on the MCP-Mod test statistic $S_{1}$. The residual-based tests also demonstrated computational advantages. Given its computational efficiency and consistent performance, the residual-based randomization test, utilizing penalized MLEs, emerged as the superior option for testing within the MCP-Mod framework. 

Our findings were particularly evident in the different data-generating scenarios originating from our trial example, see Section \ref{Sec:MainSimulation}. We can conclude that a randomization test, such as the residual-based randomization test using penalized MLEs, can be a more effective analysis option, even in the presence of a time trend. Additionally, by extending the randomization test framework to pharmacometric settings it demonstrated its feasibility and effectiveness, even when applied to continuous endpoints in Section \ref{Sec:Pharmaco}.

\subsection{Future Research Directions}

To address the issues related to complete separation, we chose Firth's method. However, alternative approaches could be explored, such as adding one success and one failure to each arm, as suggested by Clogg \citep{Clogg1991}.

While our study focused on complete randomization, random allocation, and permuted block randomization, future research should explore other randomization schemes. Various restricted randomization procedures for multi-arm trials with equal or unequal allocation ratios can be implemented using \textit{Incertus.jl} software \citep{RyeznikSverdlov2024}. More complex randomization designs, such as covariate-adaptive randomization \citep{Hu2014}, response-adaptive randomization \citep{Robertson2023}, and covariate-adjusted response-adaptive randomization \cite{ rosenberger2016randomization, RosenbergerSverdlov2008} could offer additional insights; however, to the best of our knowledge, randomization-based tests following such designs in multi-arm trial settings have not been developed yet. 

The efficacy of randomization tests could be further examined using different test statistics. We evaluated test statistics motivated by the MCP-Mod approach, but also other test statistics could be employed offering improved performance under specific conditions.
Notably, we also did not alter the modeling part of the MCP-Mod framework. Exploring randomization based inference for other dose-finding methods would also be of interest. 


Given the promising results, future research could explore the broader application of randomization tests in pharmacometric models and trial designs with additional adaptive elements. These trials often involve complex designs and high-dimensional data, where randomization-based approaches could offer robust solutions, especially in the presence of small sample sizes, unbalanced treatment groups and time trends. A natural extension would be extending our work to the longitudinal MCP-Mod framework by Buatois et al. (2021) \citep{buatois2021clrt}.

In conclusion, our study underscores the potential of randomization-based inference methods and penalized MLE for MCP-Mod. By addressing challenges such as complete separation, enhancing power while maintaining robust type-I error control, these methods offer a valuable alternative to conventional population-based approaches. Future research exploring these promising directions will further solidify the utility of randomization-based tests in Phase II studies.

\section*{Acknowledgments}

We would like to thank our colleagues Rajkumar Radhakrishnan and Astrid Jullion, who provided us with background on the trial example.

\section*{Financial disclosure}

LP: PhD-funding MRC Biostatistics Unit Core Studentship and the Cusanuswerk e.V.

\section*{Supporting information}

Code is publicly available \cite{pin2024mcpmod}.

\bibliographystyle{unsrtnat}
\bibliography{references}







\end{document}